
\input jnl
\def\bigans{b }
\message{ big or little (b/l)? }\read-1 to\answ
\ifx\answ\bigans\message{(This will come out double spaced, 12 point.}
\twelvepoint
\else\message{(This will come out single spaced, 10 point.}
\tenpoint
\fi
\def\today{{\rm\ifcase\month\or
January\or February\or March\or April\or May\or June\or
July\or August\or September\or October\or November\or December\fi
\space\number\day, \number\year}}
\def\ve{\varepsilon}
\def\L{{\cal L}}
\def\psibar{\bar{\psi}}
\def\rhobar{\bar{\rho}}
\def\betabar{\bar{\beta}}
\def\x{{\bf x}}
\def\y{{\bf y}}
\def\z{{\bf z}}
\def\ulra#1{\raise1.5ex\hbox{$\leftrightarrow$}\mkern-16.5mu #1}
\def\dslash{\not\!\!\partial}
\def\sigbar{\bar{\sigma}}
\def\sigtil{\tilde{\sigma}}

\def\epsilonbar{\bar{\epsilon}}
\def\rtn{\tilde{r}_N}
\def\rhn{\hat{r}_N}
\def\rtnb{\tilde{r}^\dagger_N}
\def\rhnb{\hat{r}^\dagger_N}
\def\rhotn{\tilde{\rho}^{(N)}}
\def\rhohn{\hat{\rho}^{(N)}}

\def\k{{\bf k}}
\def\bb{}
\def\rr{}
\def\rb{}
\def\rtob{}
\def\btob{}
\def\Vrule{\smash{\vrule height7pt depth\baselineskip}}
\def\LVrule{\smash{\vrule height1.5\baselineskip depth1pt}}
\def\Squeeze{\noalign{\vskip-.5\baselineskip}}
\def\Hrule #1{\Squeeze\multispan#1\hrulefill}
\beginparmode\oneandathirdspace
\centerline{\bf PHYSICAL STATES IN CANONICALLY QUANTIZED SUPERGRAVITY}
\vskip 1 true cm
\centerline{Sean M. Carroll\footnote{$^\dagger$}{e-mail:
  carroll@marie.mit.edu}}
\centerline{\tenit Center for Theoretical Physics,
Laboratory for Nuclear Science and Department of Physics}
\centerline{\tenit Massachusetts Institute of Technology,
Cambridge, MA 02139, USA}
\smallskip
\centerline{Daniel Z. Freedman\footnote{*}{Permanent address:
{\tenit
Department of Mathematics and Center for Theoretical Physics, Massachusetts
Institute of Technology, Cambridge, MA 02139, USA.}  e-mail:
dzf@math.mit.edu}}
\centerline{\tenit CERN, Theory Division,
CH-1211, Geneva 23, Switzerland}
\smallskip
\centerline{Miguel E.
Ortiz\footnote{$^\ddagger$}{e-mail: ortiz@mitlns.mit.edu}}
\centerline{\tenit Center for Theoretical Physics,
Laboratory for Nuclear Science and Department of Physics}
\centerline{\tenit Massachusetts Institute of Technology,
Cambridge, MA 02139, USA}
\smallskip
\centerline{and}
\smallskip
\centerline{Don N. Page\footnote{$^{\S}$}{e-mail: don@phys.ualberta.ca}}
\centerline{\tenit CIAR Cosmology Program, Theoretical Physical Institute}
\centerline{\tenit Department of Physics, University of Alberta,
Edmonton, Alberta, Canada T6G 2J1}
\vskip 1 true cm
\centerline{Submitted to Nuclear Physics B}
\vskip 1 true cm
\centerline{\bf Abstract}
We discuss the canonical quantization of $N=1$ supergravity in the
functional Schr\"o\-dinger representation. Although the form of the
supersymmetry constraints suggests that there are solutions of definite
order $n$ in the fermion fields, we show that there are no such states for
any finite $n$. For $n=0$, a simple scaling argument definitively excludes
the purely bosonic states discussed by D'Eath. For $n>0$, the argument is
based on a mode expansion of the gravitino field on the quantization
3-surface. It is thus suggested that physical states in supergravity have
infinite Grassmann number. This is confirmed for the free spin-3/2 field,
for which we find that states satisfying the gauge constraints contain an
infinite product of fermion mode operators.
\vfill

\noindent{CTP \# 2279\hfill January 1994}\par
\noindent{hep-th/9401155\hfill~}\par
\endtitlepage
\ifx\answ\bigans
\doublespace
\else
\fi

\head{\bf 1. Introduction}
\taghead{1.}

    Since supergravity is a theory with gauged space-time symmetries,
physical state functionals in the quantum theory satisfy constraint
equations that contain non-perturbative information of potential
importance [\cite{teitelboim,fradkin,d84}].
The quantum constraint equations were discussed in 1984 by
D'Eath [\cite{d84}], and he has recently [\cite{d93}] analyzed the
equations for (3+1)-dimensional supergravity to obtain surprising
information about exact solutions. This work has motivated us to study
the situation, and we present our results below.

    There are constraints for each of the gauge symmetries contained in the
theory, viz local Lorentz invariance, supersymmetry, and spacetime
diffeomorphisms. Lorentz invariance is easy to implement,
and the Dirac/Poisson
bracket structure ensures that all constraints are satisfied by a Lorentz
invariant supersymmetric state. The supersymmetry constraints are thus the
central issue. Since these constraints are homogeneous in the gravitino
field $\psi_{Ai}(\x)$, it is consistent to look for solutions involving
homogeneous functionals of order $\psi^n$. Such states may be called
Grassmann number $n$ states.

    A simple scaling argument can be applied to the supersymmetry
constraint to show that there are no bosonic $(n=0)$ states. The situation
for states of finite, non-zero Grassmann number $n$ is analyzed using a mode
decomposition of $\psi_{Ai}(\x)$ on an initial-value surface $\Sigma$.
We present an argument that there are also no states with finite $n$.
This requires explicit knowledge of the mode functions to ascertain the
linear independence of various terms in the constraint equations.
The mode equations, which we believe to be new, can be solved explicitly only
when $\Sigma$ is a flat 3-surface, when the linear independence is then easily
shown. We believe, but have no explicit proof,
that this property holds for a general surface $\Sigma$.
We therefore conclude that all physical states in canonically quantized
supergravity in a holomorphic or anti-holomorphic representation have
infinite Grassmann number.

As a check on this conclusion, we have examined the free
spin-3/2 field which has a residual abelian gauge supersymmetry and
corresponding constraints. We solve the constraints explicitly using the
mode-number basis, and we find that the most general physical state
functional contains an infinite product of Grassmann-valued mode
coefficients. Such a state may be considered to have infinite Grassmann
number in the sense described above. It is instructive to note that there are
two types of infinity present in both the holomorphic and anti-holomorphic
representations, one due the structure of the Dirac sea, and a
second due to the fixing of gauge degrees of freedom.

     Our conclusions about physical states in the free and interacting
theories are compatible with a recent analysis of 2+1 and perturbative 3+1
dimensional supergravity [\cite{nicolai}].  In the final section we
explain how our conclusions are also compatible with minisuperspace
studies, which have found bosonic physical states [\cite{mini}].
Unfortunately we disagree with
D'Eath who has claimed in [\cite{d93}] to have
exhibited exact bosonic state functionals which satisfy the constraints
in the full theory. Such states are definitively ruled out by the scaling
argument referred to above.

\head{\bf 2. Canonical quantization of N=1 supergravity}
\taghead{2.}

For Weyl spinor gravitino fields $\psi_{A\mu}$ and $\bar{\psi}_{A'\mu}$
and vierbein $E^a{}_\mu$, the covariant $N$=1 supergravity Lagrangian is
$$
\L={1\over 8\kappa^2}\ve^{\mu\nu\rho\sigma}\ve_{abcd}{E^a}_\mu{E^b}_\nu
{R^{cd}}_{\rho\sigma}-{1\over 2}\ve^{\mu\nu\rho\sigma}\left(\psibar_\mu
{E^a}_\nu\sigbar_a {D_\rho}\psi_\sigma-{D_\rho}\psibar_\mu
{E^a}_\nu\sigbar_a\psi_\sigma
\right)\ .
\eqno(2.1)
$$
Here  $\ve^{\mu\nu\rho\sigma}$ and
$\ve_{abcd}$ are Levi-Civita tensors with curved and flat indices,
respectively and $g_{\mu\nu}$ and the flat metric $\eta_{ab}$
have signature $+,-,-,-$. ${R^{cd}}_{\rho\sigma}$ is the curvature
tensor defined using a connection with torsion
${W^{ab}}_\mu={\Omega^{ab}}_\mu+{\kappa^{ab}}_\mu$, where
${\Omega^{ab}}_\mu$ is the usual torsion free connection defined by the
requirement that $D^{\Omega}_{[\mu}{E^a}_{\nu]}=0$, and
${\kappa^{ab}}_\mu$ is the contortion,
$$
{\kappa^{ab}}_\mu={i\kappa^2\over 2}{E^{a\rho}} {E^{b\sigma}}\left[
\psibar_\sigma {E^c}_\rho \sigbar_c \psi_\mu +
\psibar_\mu {E^c}_\sigma \sigbar_c \psi_\rho -
\psibar_\rho {E^c}_\mu \sigbar_c \psi_\sigma\right]\ .
\eqno(2.2)
$$
The derivative operator $D_\rho$ defined with the full connection
${W^{ab}}_\rho$ acts on spinor indices as
$$
D_\rho\psi_{A\mu}=\partial_\rho\psi_{A\mu}+{{W_A}^B}_\rho\psi_{B\mu},\qquad
D_\rho\bar{\psi}_{A'\mu}=\partial_\rho\bar{\psi}_{A'\mu}+\bar{\psi}_{B'\mu}
{\bar{W}^{B'}}_{A'\rho}\ ,
\eqno(2.3)
$$
where ${{W_A}^B}_\rho=-{1\over 4}W_{ab\rho}(\sigma^a\sigbar^b{)_A}^B$, and
${{\bar{W}^{B'}}}_{A'\rho}=-{1\over 4}
W_{ab\rho}(\sigbar^b\sigma^a{)^{B'}}_{A'}$.
The Lagrangian \(2.1) is invariant under Lorentz transformations,
general co-ordinate transformations, and the supersymmetry transformations
$$
\delta{E^a}_\mu=-{i\kappa^2\over 2}\left[\epsilonbar\sigbar^a\psi_\mu
-\psibar_\mu\sigbar^a\epsilon\right],\qquad
\delta\psi_\mu=D_\mu\epsilon,\qquad
\delta\psibar_\mu=D_\mu\epsilonbar\ .
\eqno(2.4)
$$
The canonical quantization of this theory on a spacelike 3-surface
$\Sigma$ has been developed in [\cite{d84,teitelboim,pilati}].
We summarize here only
those aspects of the formalism required to define the quantum theory and
its supersymmetry constraints. Our conventions, which differ from those of
[\cite{d84,pilati}], are summarized in Appendix A.

Let $x^i$, $i=1,2,3$, denote co-ordinates on $\Sigma$. The dynamical
variables in the Hamiltonian formalism are $e^a{}_i(\x)=E^a{}_i$, with
$a=0,1,2,3$, conjugate momenta
$p^j{}_a(\x)=\pi^i{}_a-\ve^{ijk}\psibar_j\sigbar_a\psi_k/2$ with
$\pi^i{}_a$ the usual momentum conjugate to $e^a{}_i$ in general relativity,
and fermionic variables
$\psi_{Ai}(\x)$ and $\psibar_{A'j}(\x)$. The Dirac/Poisson brackets for
\(2.1) imply that the following commutators and anti-commutators hold in
the quantum theory:
$$
\eqalign{
& \{\psi_{Ai}(\x),\psibar_{A'j}(\y)\}=-i\hbar D_{AA'ij}\delta(\x-\y)
\cr
& [{e^a}_i(\x),{p_b}^j(\y)]=i\hbar{\delta^a}_b{\delta^i}_j\delta(\x-\y)
\cr
& [{p^i}_a(\x),{p^j}_b(\y)]=-{i\hbar\over 4}\ve^{jln}\ve^{ikm}\psibar_m(\x)
\sigbar_a D_{kl}\sigbar_b\psi_n(\x)\delta(\x-\y)+{\rm H.c.}
\cr
& [{p^i}_a(\x),\psi_{Aj}(\y)]
=-{i\hbar\over 2}\ve^{ikl}D_{AB'jk}{\sigbar_a}^{B'B}
\psi_{Bl}(\x)\delta(\x-\y)
\cr
& \{\psi_{Ai}(\x),\psi_{Bj}(\y)\}=\{\psibar_{A'i}(\x),\psibar_{B'j}(\y)\}
=[e^a{}_i(\x),e^b{}_j(\y)]=0\ .
}
\eqno(2.ab)
$$
The matrix $D$ and its inverse $C$, required below, are defined by
$$
\eqalign{
& D_{AA'ij}={-i\over2\sqrt{-h}}{e^a}_j{e^b}_in^c
(\sigma_a\sigbar_b\sigma_c)_{AA'}
\cr
& C^{jiA'A}(\x)=\ve^{i j k}{e^a}_k\sigbar_a^{A'A}
\cr
& C^{ijA'A}D_{AB'jk}={\delta^{A'}}_{B'}{\delta^i}_k
\cr
& D_{AA'ij}C^{jkA'B}={\delta_A}^{B}{\delta_i}^k\ .
}
\eqno(2.19)
$$

Following [\cite{d84}], we use a holomorphic representation for the
fermionic variables in which wave functionals $F[e,\psi]$ depend on
$e^a{}_i$ and $\psi_{Ai}$, while $p^i{}_a$ and $\psibar_{A'i}$ are realized
as
$$
\hat{\psibar}_{A'j}(\x)=-i\hbar D_{AA'ij}{\delta\over\delta\psi_{Ai}(\x)}\ ,
\eqno(2.ac)
$$
and
$$
{\hat{p}_a}^i=-i\hbar{\delta\over\delta{e^a}_i(\x)}-{i\hbar\over 2}
\ve^{ijk}\psi_{Ak}(\x)D_{BA'lj}{\sigbar_a}^{A'A}{\delta\over\delta
\psi_{Bl}(\x)}\ .
\eqno(2.ac1)
$$
Alternatively, one can use a formal fermionic Fourier transform to pass to
the anti-holomorphic representation
$$
\tilde{F}[e,\psibar]=\int{\prod d\psi_i(\x)}F[e,\psi]
e^{{i\over \hbar}\int d^3\x\psibar_{A'i} C^{ijA'A} \psi_{Aj}}\ ,
\eqno(2.ad)
$$
in which $\psibar(\x)$ acts by multiplication and
$\psi\sim\delta/\delta\psibar$. A formal factor $\prod_\x{\rm Det~}C$ (see
[\cite{d84}]) has been dropped for simplicity since it plays no role in our
work. One may note that the spatial integral in the exponential in \(2.ad) is
the formally conserved fermionic charge which is easily obtained from the
Lagrangian \(2.1).

Physical state functionals must satisfy constraints arising from the gauge
invariances of \(2.1).  The most important of these are the
supersymmetry constraints, given by the field equations of
$\psi_{A0}$ and $\psibar_{A'0}$ (which appear as Lagrange multipliers in
\(2.1)).  Remarkably, all nonlinear terms in $\psi$ and $\psibar$
cancel in these constraints [\cite{d84}], which can be written in the simple
form
$$
\eqalignno{
& \bar{S}^{A'}F=\left[-\ve^{ijk}e^a{}_i{\sigbar_a}^{A'A}
(D_j\psi_{Ak})+{\hbar\kappa^2\over 2}\sigbar^{aA'A}
\psi_{Ai}{\delta\over\delta e^a{}_i}\right]F=0\ ,&(2.100)
\cr
& S^AF=\left[D_j{\delta\over\delta\psi_{Aj}}+{\hbar\kappa^2\over
2}{\delta\over\delta\psi_{Bj}}D_{BA'ji}\sigbar^{aA'A}{\delta\over\delta
e^a{}_i}\right]F=0\ .& (2.101)
}
$$
The covariant derivative $D_i$ in \(2.100) and \(2.101)
is coupled as in \(2.3) to the
spinor index of $\psi_{Ai}$, with connection
$$
\omega^{ab}_i=e^{aj}\partial_{[i}e^b{}_{j]}+n^a\partial_in^b
+e^{aj}n^bn^c\partial_je^{ci}-e^{aj}e^{bk}e_{ci}\partial_je^c{}_k
- \; a\leftrightarrow b\ ,
\eqno(2.0)
$$
where $n^a$ is the future-directed timelike unit vector
normal to $\Sigma$, uniquely determined by
$$
n_ae^a{}_i=0\qquad n^an_a=1\ .
\eqno(2.70)
$$
The connection $\omega^{ab}_i$ is defined so that $D_in^a=0$
and $D_{[i}e^a{}_{j]}=0$.
The derivative $D_i$ also couples to spatial indices with the Levi-Civita
connection intrinsic to $\Sigma$, although in most expressions this
connection is absent due to antisymmetry.
The $S^A$ constraint takes a simpler form, similar
to \(2.100), on anti-holomorphic wave-functionals [\cite{d84}].

Besides the supersymmetry constraints, there are also constraints
imposing local Lo\-rentz invariance as well as the familiar Hamiltonian
and diffeomorphism constraints. The Lorentz constraint simply implies that
$F[e,\psi]$ must involve variables $e^a{}_i(\x)$ and $\psi_{Aj}(\x)$ in
manifestly local Lorentz invariant combinations, and is thus
straightforward to satisfy.  In principle,
any Lorentz invariant state $F[e,\psi]$ which satisfies the $S$ and
$\bar{S}$ constraints also satisfies the Hamiltonian and diffeomorphism
constraints ${\cal H}_0$ and ${\cal H}_i$, since
$$
\left\{S^A(\x),\bar{S}^{A'}(\y)
\right\}F[e,\psi]={1\over 2}\hbar\kappa^2\left[n_a
(\x){\cal H}(\x)+ e_a{}^i(\x){\cal H}_i(\x)\right]
\sigbar_a{}^{A'A}\delta(\x-\y)F[e,\psi]\ .
\eqno(2.dd)
$$
Thus any normalizable Lorentz invariant
solution of \(2.100) and \(2.101) is an exact state
of quantum supergravity\footnote{$^1$}{The definition of a norm is a difficult
issue in any gravitational theory. Further, to our knowledge, \(2.dd) has
been established only formally, that is without complete attention to
operator ordering. See [\cite{nicolai2}] for a discussion of operator
ordering in 2+1 dimensional supergravity.},
and it is for this reason that the proposals of [\cite{d93}] must be
studied critically.

\head{\bf 3. Bosonic States}
\taghead{3.}

Let us consider Lorentz invariant states in the holomorphic representation.
Since the supersymmetry constraints \(2.100) and \(2.101)
are homogeneous in $\psi$, it is convenient to expand
an arbitrary state in a power series in the Grassmann fermion fields:
$$
F[e,\psi]=\sum_n F^{(n)}[e,\psi]\ ,
\eqno(5.ac)
$$
where
$$
\eqalign{
& F^{(0)}[e,\psi]=F^{(0)}[e]
\cr
& F^{(2)}[e,\psi]=\int d\x_1 d\x_2 K^{(2)}[e;\x_1,\x_2)^{ij}_{AB}
\psi^A_i(\x_1)\psi^B_j(\x_2)
}
\eqno(b)
$$
and so on. Note that odd $n$ states need not be considered because they are
not local Lorentz invariant. We shall refer to $F^{(n)}$ as a state of
Grassmann number $n$. $n$ is the eigenvalue of the Weyl charge (as in the
integrand in the exponential in \(2.ad)) on $F^{(n)}$.
The constraint equations \(2.100) and \(2.101) must be satisfied independently
by each term $F^{(n)}[e,\psi]$.

We shall consider the case where $F[e,\psi]\equiv F^{(0)}[e]$ is a
functional only of the ${e^a}_i$, that is $\delta F^{(0)}/\delta\psi_{Ai}=0$
(we shall refer to states of this type as ``bosonic states''). In this
case, any Lorentz invariant state satisfying $\bar{S}^{A'}F^{(0)}=0$
automatically satisfies all other constraints, since $S^AF^{(0)}$ vanishes
identically. A solution of this type
was sought in [\cite{d93}].
We now show that no
such solution can exist.

The supersymmetry constraint can be written as [\cite{d93}]
$$
\left[F^{(0)}\right]^{-1}\bar{S}F^{(0)}\equiv
-\ve^{i j k}{e^a}_i\sigbar_a (D_j\psi_k)+{\hbar\kappa^2\over
2}\psi_i\sigbar^a{\delta(\ln F^{(0)}[e])\over\delta{e^a}_i}=0  \ .
\eqno(3.9)
$$
In this section we shall establish our contradiction by using an integrated
form of \(3.9) with an arbitrary continuous, commuting, spinor test function
$\bar{\epsilon}(\x)$:
$$
\int d^3\x \bar{\epsilon}(\x)\left[
-\ve^{i j k}{e^a}_i(\x)\sigbar_a (D_j\psi_k(\x))+{\hbar\kappa^2\over
2}\psi_i(\x)\sigbar^a{\delta(\ln F^{(0)}[e])
\over\delta{e^a}_i(\x)}\right]=0\ ,
\eqno(3.10)
$$
for all $\bar{\epsilon}(\x)$, ${e^a}_i(\x)$, and $\psi_{k}(\x)$.
We shall demonstrate that there are no solutions to this version of \(3.9).
Let the integral in Eq. \(3.10) be $I$, and let $I'=I+\Delta I$ be the
integral when $\bar{\epsilon}(\x)$ is replaced by
$\bar{\epsilon}(\x)e^{-\phi(\x)}$ and $\psi_{i}(\x)$ is replaced by
$\psi_{i}(\x)e^{\phi(\x)}$, where $\phi(\x)$ is a scalar function.
Since $\bar{\epsilon}(\x)\psi_{i}(\x)$
is unchanged, the second term (with the
functional derivative) cancels in the difference between $I'$ and $I$, so
that
$$
\Delta I=-\int d^3\x
\ve^{ijk}{e^a}_i(\x)\bar{\epsilon}(\x)\sigbar_a\psi_k(\x)
\partial_j\phi(\x)=-\int \omega(\x)\wedge d\phi(\x)\ ,
\eqno(3.a)
$$
with the two-form
$$
\omega(\x)=\left(\psi_{i}(\x)dx^i\right)\wedge
\left({e^a}_j(\x)\bar{\epsilon}(\x)\sigbar_a dx^j\right)\ .
\eqno(3.b)
$$
Notice that $\Delta I$ is independent of the state $F^{(0)}[e]$.
Clearly, it is possible to choose the arbitrary fields
$\bar\epsilon(\x)$, $\phi(\x)$, $e^a{}_i(\x)$ and $\psi_k(\x)$ such that
(3.5) is nonvanishing.

For example, consider the case of a three-torus with each
of the three $x^i$ running from 0 to $2\pi$, and let
${e^a}_i=\delta^a_i$, $\psi_{Ak}=\delta_{A1}\delta_{2k}$,
$\bar{\epsilon}_{A'}=-\delta_{A'1}f(\x)$, $f=\cos x^1$,
$\phi=\sin x^1$. Then
$$
\Delta I=\int d^3x f(\x) \partial_1\phi(\x)=\int d^3x\cos^2(x^1)
=4\pi^3\neq 0\ .
\eqno(3.c)
$$
Clearly, if $\Delta I \neq 0$, we cannot have both $I=0$ and $I'=0$,
so \(3.10) cannot be satisfied for all
bosonic $\bar{\epsilon}(\x)$, ${e^a}_{i}(\x)$ and $ \psi_{k}(\x)$.

A precisely analogous argument applies to the $S^A$ constraint
applied to an anti-holomorphic bosonic wave functional,
$ \tilde{F}^{(0)}[{e^a}_i(\x),\psibar_{i}(\x)]\equiv\tilde{F}^{(0)}
[{e^a}_i]$ independent of $\psibar_{i}(\x)$. We conclude that bosonic
wave functionals of either type are inconsistent with the supersymmetry
constraints, contrary to the claims in [\cite{d93}].

This result suggests that there may also be no
states involving a finite nonzero number of fermion fields; that is,
that all quantum wave functionals of $N=1$ supergravity
with a finite (zero or nonzero) Grassmann number would be
inconsistent with the supersymmetry constraints. We shall return to this
question in Sec.~5. In the next section we look at the free spin-3/2 field,
and show that in that case all wave functionals solving the gauge constraints
in the appropriate representation necessarily contain an infinite number of
fermion fields.

\head{\bf 4. The free spin-3/2 field}
\taghead{4.}

An understanding of the structure of physical states of the free gravitino
is a useful prelude to the interacting theory of supergravity. We will
discuss the free field from a point of view which resembles the treatment
of supergravity in the previous section and in [\cite{d84}].  The result,
that wave functionals which satisfy the constraints must have infinite
Grassmann number, suggests that we should anticipate a similar structure in
the full theory. In this flat space example, $g_{\mu\nu}={\rm
diag}(+1,-1,-1,-1)$, $\sigma_\mu=(1,\sigma_i)$ and $\varepsilon^{0123}=1$,
and for simplicity we drop the distinction between primed and unprimed
spinor indices, using greek letters instead. A perturbative quantization of
the free spin-3/2 field has previously  been given in [\cite{dandas}], and
its canonical structure is discussed in [\cite{pilati}] and [\cite{senja}].

\subhead{\bf 4.1 Constraints}

The free Lagrangian
$$
\L=-{1\over
2}\ve^{\mu\nu\rho\sigma}\psibar_\mu\sigma_\nu\ulra{\partial_\rho}
\psi_\sigma
\eqno(4.1)
$$
is gauge invariant under $\delta\psi_\mu=\partial_\mu\epsilon(\x)$, with
$\epsilon(\x)$ an arbitrary two-component spinor.
The canonical procedure [\cite{d84,pilati}] leads to the equal time
anti-commutation rules
$$
\left\{\psi_{i\alpha}(\x),\psibar_{j\beta}(\y)\right\}
={\hbar\over
2}\sigma_j\sigma_i
\delta(\x-\y)\equiv -i\hbar D_{\alpha\beta ij}\delta(\x-\y)\ ,
\eqno(4.1a)
$$
where $i,j,k=1,2,3$ and these spatial indices are raised and lowered by a
metric $h_{ij}=-\delta_{ij}$.  In this section we use Greek
letters $\alpha, \beta$ to denote spinor indices; thus,
$D_{\alpha\beta ij}=i\sigma_j\sigma_i/2$ is the
flat space version of $D_{AA'ij}$ defined in \(2.19). $\psi_0$ and $\psibar_0$
are Lagrange multipliers, and their constraints generate time-independent
gauge transformations.

In the holomorphic representation, a state of the system has a wave
functional $F[\psi]$ with Grassmann variables $\psi_{i\alpha}(\x)$, and the
operator $\psibar_{j\beta}(\y)$ is represented by
$$
\hat{\psibar}_{j\beta}(\y) F[\psi]= -i\hbar D_{ij\alpha\beta}{\delta\over
\delta\psi_{i\alpha}}F[\psi]\ .
\eqno(4.2)
$$
Equivalently, one may use the anti-holomorphic representation, obtained by
the fermionic Fourier transformation
$$
\tilde{F}[\psibar]=\int(\prod
d\psi_{i\alpha})\left[{i\over\hbar}
\exp\int\psibar_{i\alpha}(\y){C^{ij}}_{\alpha\beta}
\psi_{j\beta}(\y)d^3\y\right] F[\psi]\ ,
\eqno(4.3)
$$
with ${C^{ji}}_{\alpha\beta}=\ve^{ijk}{\sigma_k}_{\alpha\beta}$,
and ${C^{ji}}_{\alpha\beta}D_{ik\beta\gamma}=\delta^j_k\delta_{\alpha\gamma}$.
In the anti-holomorphic representation $\psi$ acts as
$$
\hat{\psi}_{i\alpha}\tilde{F}[\psibar]=-i\hbar D_{ij\alpha\beta}
{\delta\over\delta \psibar_{j\beta}}\tilde{F}[\psibar]\ .
\eqno(4.4)
$$
{}From now on we shall focus on the holomorphic representation.

Physical states satisfy constraints which may be obtained as the $\kappa\to
0$ limit of \(2.100) and \(2.101)
or as the $\psibar_0$ and $\psi_0$ field equations from
\(4.1). These constraints are
$$
\bar{S}_\alpha(\x)F[\psi]=-\ve^{ijk}\sigma_{i\alpha\beta}\partial_j
\psi_{k\beta}(\x)F[\psi]=0\ ,
\eqno(4.5)
$$
and
$$
S_\beta
F[\psi]=-\ve^{ijk}\partial_j\psibar_{k\alpha}(\x)\sigma_{i\alpha\beta}F[\psi]
=i\partial_j{\delta\over{\delta\psi_{j\beta}}}F[\psi]=0\ .
\eqno(4.6)
$$
Note that the constraints satisfy
$$
\{\bar{S}_\alpha(\x),S_\beta(\y)\}=0\ ,
\eqno(4.6a)
$$
in distinction to supergravity where a combination of the
Hamiltonian and diffeomorphism constraints appears on the right side of Eq.
\(2.dd)).

\subhead{\bf 4.2 Mode Decomposition}

We shall now proceed to find the general solution of these constraints.
As the first step in this process we obtain a set of mode equations whose
solutions give a complete orthonormal set of functions suitable for
expansion of the initial data for $\psi_{i\alpha}(\x)$. The initial-value
surface $\Sigma$ will be taken to be $R^3$, but the extension to a
3-torus is immediate. Later we will generalize the mode equations to a
3-manifold with arbitrary metric, embedded in a 3+1 dimensional spacetime.
This will give a set of mode functions which can be applied to study the
constraints in supergravity.

We decompose $\psi_i(\x)$ into spin-3/2 and spin-1/2 parts
$$
\psi_{i\alpha}(\x)=\rho_{i\alpha}(\x)+\sigma_{i\alpha\beta}
\beta_\beta(\x)\ ,
\eqno(4.7)
$$
where $\rho_i(\x)$ is a spinor vector constrained to satisfy
$$
{\sigma^i}\rho_{i}=0\ .
\eqno(4.8)
$$
The $\rho_i$ and $\beta$ parts of $\psi_i$ may be projected out by
$$
\rho_i(\x)=(\delta_{ij}-{1\over 3}\sigma_i\sigma_j)\psi^j(\x),\qquad
\beta(\x)=-{1\over 3}\sigma_i\psi^i(\x)\ .
\eqno(4.9)
$$
Under space-dependent gauge transformations, we have
$$
\delta\rho_i=(\partial_i+{1\over
3}\sigma_i\sigma^j\partial_j)\epsilon\qquad
\delta\beta=-{1\over 3}\sigma^i\partial_i\epsilon\ .
\eqno(4.?)
$$

The matrix $-i\hbar
D_{ij\alpha\beta}$ in \(4.2) has eigenvalue $\hbar$ on $\rho_i(\x)$ and
eigenvalue $-\hbar/2$ on $\sigma_i\beta(\x)$. One can use this to deduce that
$$
\left\{\rho_{i\alpha}(\x),\rhobar_{j\beta}(\y)\right\}=
\hbar\left(\delta_{ij}\delta_{\alpha\beta}
-{1\over 3}\left(\sigma_i\sigma_j\right)_{\alpha\beta}\right)\delta(\x-\y)\ ,
\eqno(4.10)
$$
and
$$
\left\{\beta_{\alpha}(\x),\betabar_{\beta}(\y)\right\}=
-{\hbar\over 6}\delta_{\alpha\beta}\delta(\x-\y)\ ,
\eqno(4.11)
$$
with the cross brackets vanishing. It also follows that
$$
\tilde{F}[\rhobar,\betabar]=\int(\prod d\rho_{i\alpha}d\beta_\beta)
\exp\int\left[-{1\over\hbar}
d^3\y \left(-\rhobar_i\rho_j\delta^{ij}+6\betabar\beta\right)\right]
F[\rho,\beta]\ .
\eqno(4.12)
$$

When the decomposition \(4.7) is inserted into \(4.5), we find that the
$\bar{S}$ constraint reads
$$
(\partial_i\rho^i+2\sigma^i\partial_i\beta)F[\rho,\beta]=0\ .
\eqno(4.15)
$$
The corresponding form of the $S$ constraint is
$$
(\partial_i\rhobar^i+2\partial_i\betabar\sigma^i)F[\rho,\beta]=0\ .
\eqno(4.16)
$$
But \(4.10) and \(4.11) tell us that, in the holomorphic representation,
$\rhobar_i(\x)$ and $\betabar(\x)$ are realized by
$$
\rhobar_i=\hbar{\delta\over\delta\rho_j}\left(\delta_{ji}-{1\over
3}\sigma_j\sigma_i\right),\qquad
\betabar=-{\hbar\over 6}{\delta\over\delta\beta}\ ,
\eqno(4.17)
$$
so that \(4.16) is equivalent to
$$
\left[\partial_i{\delta\over\delta\rho_j}\left(\delta^i_j+{1\over
3}\sigma_j\sigma^i\right)+{1\over 3}
\partial_i{\delta\over\delta\beta}\sigma^i\right]
F[\rho,\beta]=0\ .
\eqno(4.18)
$$

Now let $\dslash=-\sigma^i\partial_i$ denote the Dirac operator on $R^3$.
Then the Dirac equation
$$
i\dslash\beta^{(N)}=\lambda_N\beta^{(N)}
\eqno(4.19)
$$
gives a complete set of modes for $\beta(\x)$, specifically two modes for
each momentum ${\bf k}$, with $\lambda=\pm |{\bf k}|$.
The right mode equation for $\rho_i(\x)$ is somewhat trickier.
The most general first order equation is
$$
i\left[\dslash\rho_i^{(n)}+\alpha\sigma_i\partial_j\rho^{j(n)}\right]=
\mu_n\rho_i^{(n)}\ ,
\eqno(4.20)
$$
with real parameter $\alpha$. Upon contraction with $\sigma^i$ and use of
\(4.8), one finds that
$$
\left(\alpha-{2\over 3}\right)\partial_i\rho^{i(n)}=0\ .
\eqno(4.21)
$$
We cannot accept an additional $\partial\cdot\rho=0$ constraint on
the $\rho_i^{(n)} $ for two reasons. It does not allow the
implementation of the
gauge transformations \(4.?), and it cuts down on the number of
solutions of \(4.20). Namely, there are only two solutions of \(4.20) for
each momentum ${\bf k}$, when $\alpha\ne 2/3$, whereas we clearly need four
solutions to represent a general configuration for $\rho_i(\x)$. Therefore
we choose $\alpha=2/3$, and the mode equation
$$
i\left[\dslash\rho_i^{(n)}+{2\over
3}\sigma_i\partial_j\rho^{j(n)}\right]=\mu_n\rho^{(n)}_i\ .
\eqno(4.22)
$$
It is easy to show that the operator on the left side is self-adjoint with
respect to the natural scalar product
$$
\int d^3\x \rhobar^i(\x)\delta_{ij}\rho^j(\x)\ ,
\eqno(4.23)
$$
for fields satisfying $\sigma^i\rho_i=0$.

One may distinguish two classes of solutions to equation \(4.22). There are
``pure gauges" which are related to the $\beta^{(N)}$ modes by
$$
\tilde{\rho}^{(N)}_i={i\over\lambda_N}\sqrt{3\over 2}\left(\partial_i-{1\over
3}\sigma_i\dslash\right)\beta^{(N)}\ ,
\eqno(4.24)
$$
and have eigenvalues $\tilde{\mu}_N=\lambda_N/3$, with two such modes
for each $\k$. There is also a set of modes $\hat{\rho}^{(N)}_i$,
again two for each $\k$, which
are orthogonal to all pure gauges and thus satisfy
$\partial_i\hat{\rho}^{i(N)}=0$ in addition to \(4.8) and
\(4.22).\footnote{$^2$}{For every eigenvalue, there are two $\tilde\rho$
modes, two $\hat\rho$ modes, and two $\beta$ modes.  We therefore use
uppercase superscripts $(N)$ to index all of these sets; lowercase
superscripts $(n)$ are used to index the entire collection of $\rho$
modes.}  It is the
$\hat{\rho}^{(N)}_i$ modes which describe the physical helicity $\pm3/2$
excitations of the system. For completeness we present the full set of
normalized solutions of our mode equations in terms of helicity spinors
$\chi^\pm({\bf k})$ and circular polarization vectors $\ve^a_i(\k)$, with
$a=0,\pm$
$$
\eqalign{
& \tilde{\rho}^{(\pm,\k)}_i(\x)=e^{-i\k\cdot\x}\left(\mp
{2\over\sqrt{3}}\chi^{\mp}(\k)\ve^\pm_i(\k)+{1\over
\sqrt{3}}\chi^\pm(\k)\ve^0_i(\k)\right),\qquad \mu=\pm{1\over 3}|\k|
\cr
& \hat{\rho}^{(\pm,\k)}_i(\x)=
e^{-i\k\cdot\x}\chi^{\pm}(\k)\ve^\pm_i(\k),\qquad \mu=\pm |\k|
\cr
& \beta^{(\pm,\k)}(\x)=e^{-i\k\cdot\x}\chi^\pm(\k),\qquad \lambda=\pm |\k|
}
\eqno(4.25)
$$
where if, for example, $\k=(0,0,k)$, then
$$
\ve^\pm_i={1\over\sqrt{2}}(1,\pm i,0),\qquad
\ve^0_i=(0,0,1),\qquad
\chi^+={1\choose 0},\qquad
\chi^-={0\choose 1}\ .
\eqno(4.25a)
$$
For general momenta, $\ve_i(\k)$ and $\chi(\k)$ are obtained from these
expressions by rotations. Note that the modes are normalized so that
$$
\eqalign{
&\sum_n\rho^{(n)}_i(\x)\rhobar^{(n)}_j(\y)=\left(\delta_{ij}-{1\over
3}\sigma_i\sigma_j\right)\delta(\x-\y)\cr
& \sum_n\beta^{(N)}(\x)\betabar^{(N)}(\y)=\delta(\x-\y)\ ,
}
\eqno(4.25b)
$$
and
$$
\eqalign{
& \int d^3\x\rhobar^{(n)i}(\x)\delta_{ij}\rho^{(m)j}(\x)=\delta_{mn}\cr
& \int d^3\x \betabar^{(N)}(\x)\beta^{(M)}(\x)=\delta_{MN}\ .
}
\eqno(4.25c)
$$

On the initial value surface the $\rho_i$ and $\beta$ parts of $\psi_i$ can
be expanded in the mode system as
$$
\eqalign{
& \rho_i(\x)=\sum_N\left[\rtn\rhotn_i(\x)+\rhn\rhohn_i(\x)\right]\ ,
\cr
& \beta(\x)=\sum_N b_N\beta^{(N)}(\x)}\ ,
\eqno(4.26)
$$
where $\rtn$, $\rhn$ and $b_N$ are Grassmann numbers, while
$$
\eqalign{
&{\delta\over\delta\rho_j(\x)}=-\sum_N\left[\tilde{\bar{\rho}}^{j(N)}
(\x){\delta\over\delta
\rtn}+\hat{\bar{\rho}}^{j(N)}(\x){\delta\over\delta\rhn}\right]
\cr
& {\delta\over\delta\beta(\x)}=\sum_N\bar{\beta}^{(N)}(\x)
{\delta\over\delta b_N}\ .
}
\eqno(4.27)
$$
In this mode basis we can replace the wave functional $F[\psi]$ by
$F[\rtn,\rhn,b_N]$. This representation is, as we shall see below,
equivalent to a Fock space representation.

Finally, we remark that in [\cite{d93}],
D'Eath presented a decomposition of $\psi_{i\alpha}(\x)$ in terms of
a spinor $\beta_{\alpha}(\x)$ and a totally symmetric tri-spinor
$\gamma_{\alpha\beta\gamma}(\x)$. He then specified mode equations for
$\beta$ and $\gamma$ for a general initial-value surface $\Sigma$. The
equations are easy to solve for flat $\Sigma$, and results can be compared
with ours. One finds that his $\gamma^{(n)}_{\alpha\beta\gamma}$ modes
satisfy a constraint analogous to \(4.21) and that there are only two
$\gamma^{(n)}$ modes for each momentum {\bf k}. So the mode equations
presented in [\cite{d93}]
do not lead to a general solution of the initial value
problem for $\psi_{i\alpha}(\x)$, although
it is possible that they can be
modified to eliminate the unwanted constraint.
Certainly our decomposition \(4.9) and mode equations \(4.19) and \(4.22) are
not unique, but any correct choice must produce six independent modes for
every momentum {\bf k}.

\subhead{\bf 4.3 Physical states}

Using the relation \(4.24) between $\beta$ and $\tilde{\rho}$ modes,
and mode orthogonality, one can see that
the constraints \(4.15) and \(4.18) become
$$
\left[\rtn-\sqrt{6}b_N\right]F=0\ ,\quad\forall N\neq 0\ ,
\eqno(4.28)
$$
and
$$
\left[{\delta\over\delta\rtn}+{1\over\sqrt{6
}}{\delta\over\delta
b_N}\right]F=0\ ,\quad\forall N\neq 0\ ,
\eqno(4.29)
$$
where the modes with eigenvalue $\lambda_0\equiv 0$ and $\tilde\mu_0
\equiv 0$ drop out of the constraint. The most general physical state
functional is then
$$
F=\left[\prod_M\left({{\tilde{r}_M}\over\sqrt{6}}-b_M\right)\right]
f[\rhn,\tilde{r}_0,b_0]\ ,
\eqno(4.30)
$$
where the infinite product extends over all non-zero modes and
$f[\rhn,\tilde{r}_0,b_0]$ is an arbitrary function of the indicated
Grassmann variables. Here $\tilde{r}_0$ and $b_0$ denote the Grassmann
coefficients corresponding to zero modes.
In what follows, we shall ignore the dependence on these
zero modes, since
in flat space free field theory, they are usually thrown away.
On the torus these modes are of course present, and may be of importance.

In order to make contact with standard Fock space
quantization methods, it is now
useful to convert $r_n$ and $\bar{r}_n$ into conventional creation and
annihilation operators. We define
$$
c_\k=\hat{r}_{(+,\k)},\qquad d^\dagger_\k=\hat{r}_{(-,\k)}\ ,\qquad
d_\k=\hat{\bar{r}}_{(-,\k)},\qquad
c_\k^\dagger=\hat{\bar{r}}_{(+,\k)}\ .
\eqno(4.1aa)
$$
where the dagger is just notation, since we have not yet defined an inner
product on physical states. They obey the anti-commutation relations
$$
\{c_\k,c_{\k'}^\dagger\}=\delta_{\k\k'}\ ,\qquad
\{d_\k,d_{\k'}^\dagger\}=\delta_{\k\k'}\ ,
\eqno(4.1ab)
$$
with all others zero. Clearly, in our chosen representation,
$$
d_\k={\delta\over\delta d_\k^\dagger}\ ,\qquad c_\k^\dagger={\delta\over\delta
c_\k}\ .
\eqno(4.1ac)
$$
We may now compute the Hamiltonian. In terms
of the fields $\psi(\x)$ and $\psibar(\x)$, this takes the form
$$
H=\int d^3\x\left({1\over 2}\ve^{ijk}\left(\partial_i\psibar_k
\right)\psi_j-{1\over 2}\ve^{ijk}
\psibar_k\partial_i\psi_j+\psibar_0\left[\ve^{ijk}\sigma_i\partial_j\psi_k
\right]-\psi_0\left[\ve^{ijk}\partial_j\psibar_k\sigma_i\right]\right)
\eqno(4.1ad)
$$
In terms of the operators $b_N$, $c_N$ and $d_N$ and their conjugates,
it becomes
$$
H=\sum_{\pm,\k}|\k|\left[
{\delta\over\delta c_{\k}}c_\k
+d_\k^\dagger{\delta\over\delta d_\k^\dagger}
\right]\ ,
\eqno(4.1ae)
$$
up to a zero-point constant and terms proportional to the constraints.
At this stage, we may effectively ignore the infinite product in \(4.30),
since it commutes with the Hamiltonian, and regard a state as being defined
by its $f[\rhn]$ part. In terms of the new variables, $f$ is a function of
the $c_\k$ and $d^\dagger_\k$.

We may now construct the usual Fock states, starting with the vacuum state
state $f_0[c_\k,d^\dagger_\k]$, which is
$$
f_0=\prod_\k c_\k\ ,
\eqno(4.1af)
$$
so that $c_\k f_0=0$ and $\delta f_0/\delta d^\dagger_\k=0$,
$\forall \k$. A state with a single excitation is created
by $c_\k^\dagger\equiv \delta/\delta c_\k$
as $f_{(+,\k)}= \delta f_0/\delta c_\k$, or by
$d_\k^\dagger$ as $f_{(-,\k)}=d_\k^\dagger f_0$,
showing the Fock character of the state space.
The infinite product of $c_\k$'s in the vacuum state
represents the filling of the
Dirac sea. The ``bare'' vacuum, $f_0=1$, is annihilated by the
$c^\dagger_\k$, and has negative energy excitations
created by multiplication by $c_\k$, while the true vacuum, with its infinite
product of $c_\k$'s, has all these negative energy
states filled. A similar result applies in the anti-holomorphic
representation, where the vacuum state $\tilde{f}_0[c^\dagger_\k,d_\k]$
has an infinite product of $d_\k$'s.

Our main conclusion in this section is that all physical states must have
the form \(4.30), and therefore have infinite Grassmann number. This
statement is independent of the problem of defining an inner product on the
space of states, but we now discuss this issue briefly.
The natural inner product appropriate
for holomorphic quantization is [\cite{d84}, \cite{fadslav}]
$$
\langle F\vert G\rangle
 =\int(\prod d\psi d\psibar)
\tilde{F}[\psibar]\exp\left[-{i\over\hbar}
\int d^3\y \psibar C\psi\right] G[\psi]\ .
\eqno(4.31)
$$
In the mode number basis this becomes
$$
\langle F|G\rangle=\int\prod_N d\rhn
d\rtn db_N d\rtnb d\bar{b}_N d\rhnb
\tilde{F}\exp\left[{1\over\hbar}
\sum_N \left(\rhnb\rhn+\rtnb\rtn-6\bar{b}_N b_N
\right)\right]G\ .
\eqno(4.32)
$$
The minus sign in front of the $\bar{b}_N b_N$ term means that the
metric is indefinite, as required by \(4.11).

Direct computation reveals that this inner product vanishes on
all physical states, due to the prefactor in \(4.30), its
anti-holomorphic analogue, and the form of the exponential
in \(4.32).  This is expected in
a constrained system, since the inner product involves
integration over gauge degrees of freedom on which physical
states do not depend.  In
bosonic theories, the inner product is infinite on all states unless this
overcounting is eliminated [\cite{cs}]. In the case at hand, \(4.32)
includes Berezin integrations over all unphysical modes, which
gives zero rather than infinity.

We can remedy this by
introducing a gauge-fixing condition for space dependent gauge
transformations into the functional integrals \(4.31) or
\(4.32), in the form of a delta function [\cite{cs}].
There are many choices for a gauge
fixing condition --- for example, we might wish to choose the condition
$\partial_i\rho^i-2\sigma^i\partial_i\beta=0$ and its conjugate,
which have non-vanishing Dirac brackets with the constraints.
A delta function imposing
these conditions inserts an infinite product of $(\tilde{r}_N+\sqrt{6}b_N)$
and their hermitian conjugates into \(4.32). It is straightforward to see that
the new inner product with this delta function included
has the correct norm on the Fock states. For example,
the vacuum state $F_0=[\prod_M(\tilde{r}_M/\sqrt{6}-b_M)]f_0$ has norm 1.

\head{\bf 5. Physical states in N=1 supergravity}
\taghead{5.}

In this section, we shall discuss the fermionic dependence of states
satisfying $\bar{S}^{A'}F=0$, where $F[e,\psi]$ is not purely
bosonic. We shall examine separately states
$F^{(n)}[e,\psi]$, with Grassmann number $n$, according to the
decomposition introduced in Sec. 3.
We write the constraint \(2.100) acting on $F^{(n)}$ as
$$
\epsilon^{ijk}e^a{}_i(\x)\bar\sigma_a[D_j\psi_k(\x)]F^{(n)}[e,\psi]
  -{{\hbar\kappa^2}\over 2}\bar\sigma^a\psi_i(\x){{\delta F^{(n)}}[e,\psi]
  \over
  {\delta {e^a}_i(\x)}}=0\ .
\eqno(c)
$$
States representing an odd number of fermions do not satisfy the
Lorentz constraint, so we shall only consider states with an even
Grassmann number. The aim of this section is to
show that no finite-$n$ state can satisfy \(c), a result
consistent with our findings for the free spin-3/2 field.

\subhead{\bf 5.1 Mode expansion in curved space}

In order to discuss states of definite Grassmann number, it is convenient to
introduce a decomposition of the fields $\psi_i$ and $\psibar_i$ in a
similar way to that used in flat space in section 4. However, in curved
space it is useful to define
$$
\sigtil_{iA}{}^B\equiv
n^ae^b{}_i(\sigma_a\sigbar_b)_A{}^B
\eqno(5.aac)
$$
to replace the flat space sigma matrices. Note that $\sigtil_i(\x)$
depends on the $e^a{}_i(\x)$.
In flat space, where $e^a{}_i=\delta^a{}_i$ and $n^a=(1,0,0,0)$,
$\sigtil_i$ is an ordinary sigma matrix. In general, the $\sigtil_i$
share many identities with the flat space sigma-matrices, the most useful
being:
$$
\sigtil_i\sigtil_j=-h_{ij}+i\sqrt{-h}\ve_{ijk}\sigtil^k\ ,\qquad
\sigtil^i\sigtil^j=-h^{ij}+{i\over\sqrt{-h}}\ve^{ijk}\sigtil_k\ .
\eqno(5.ad)
$$
Using $\sigtil_i$, we can write
$$
\psi_k=\rho_k+\sigtil_k\beta\ ,
\eqno(5.1)
$$
with $\sigtil^k\rho_k=0$. The space of functions $\rho_k(\x)$ and
$\beta(\x)$
are spanned by solutions of the curved space versions of \(4.19) and
\(4.20),
$$
i\sigtil^iD_i\rho_j^{(n)} - {2\over 3}i
\sigtil_jD_i\rho^{i(n)}=\mu_n\rho_j^{(n)}
\eqno(5.2)$$
and
$$
i\sigtil^i D_i\beta^{(M)} = \lambda_M \beta^{(M)}\ .
\eqno(5.3)
$$
Here the modes $\rho^{(m)}_i$ and $\beta^{(M)}$ are normalized so that
$$
\eqalign{
& \sum_n
\rho^{(n)}_{Ai}(\x)\rhobar^{(n)}_{A'j}(\y)=
-{1\over\sqrt{-h}}\left(h_{ij}\delta_A{}^C-{1\over
3}(\sigtil_i\sigtil_j)_A{}^C\right)\sigma^a_{CA'}n_a\delta^{(3)}(\x-\y)
\cr
& \sum_M
\beta^{(M)}_{A}(\x)\betabar^{(M)}_{A'}(\y)=
{1\over\sqrt{-h}}\sigma^a_{AA'}n_a\delta^{(3)}(\x-\y)\ ,
}
\eqno(5.ae)
$$
and
$$
\eqalign{
& -\int
d^3\x\sqrt{-h}\rhobar^{(n)}_{A'i}(\x)h^{ij}(\x)\rho^{(m)}_{Aj}(\x)
{\sigbar_a}^{A'A}n^a=\delta_{mn}
\cr
& \int d^3\x\sqrt{-h}\betabar^{(N)}_{A'}(\x)\beta^{(M)}_A(\x)
{\sigbar_a}^{A'A}n^a=\delta_{MN}\ .
}
\eqno(5.af)
$$
Note that the first mode equation implies the relation
$$
{i\over 3}\sigtil^k D_k(D_i\rho^{i(n)})+{3i\over 2}
R_{ik}\sigtil^k\rho^{i(n)}=\mu_n(D_i\rho^{i(n)})\ ,
\eqno(5.4)
$$
so that $D_i\rho^i$ satisfies a $\beta$-type equation
only on a surface of constant curvature $R_{ij}=Kh_{ij}$.

Having defined the quantity $\sigtil_i$, it is useful to rewrite the
constraint equation \(c), multiplied by $n^a\sigma_a$, as
$$
\ve^{ijk}\sigtil_i(\x)D_j\psi_k(\x) F^{(n)}[e,\psi]-{\hbar\kappa^2\over
2}n_a(\x)\sigma^a\sigbar^b\psi_i(\x)
{\delta F^{(n)}[e,\psi]\over\delta e^b{}_i(\x)}=0\ .
\eqno(5.c)
$$
Inserting the decomposition \(5.1)
for $\psi_k$ into the constraint \(5.c), we
find that the coefficient of $F^{(n)}$ in the first term of the constraint
becomes
$$
\ve^{ijk}\sigtil_iD_j\psi_k=
2i\sqrt{-h}\sigtil^jD_j\beta-{i}{\sqrt{-h}}D_j\rho^j\ .
\eqno(5.5)
$$

A general gravitino field may be written in terms of the modes
$\rho^{(m)}_i$ and $\beta^{(M)}$ as
$$
\psi_i(\x)=\sum_mr_m\rho^{(m)}_i(\x)
+\sum_Mb_M\sigtil_i\beta^{(M)}(\x)\ .
\eqno(5.ag)
$$
This allows us to write each $F^{(n)}$ in terms of modes; for example,
$F^{(2)}$ of \(b) becomes
$$
F^{(2)}[e,r,b]=r_{m}r_{n}K^{\rr}_{m n}[e]+
  r_{m}b_{N}K^{\rb}_{m N}[e]+
  b_{M}b_{N}K^{\bb}_{M N}[e]\ ,
\eqno(f)
$$
where
$$
\eqalign{
2K^{\rr}_{mn}[e]&=\int d\x d\y\ K^{(2)}
  [e,\x,\y)^{ij}\left(\rho^{(m)}_i(\x)\rho^{(n)}_j(\y)
  - m\leftrightarrow n\right)
\cr
K^{\rb}_{mN}[e]&=\int d\x d\y\ K^{(2)}
  [e,\x,\y)^{ij}\left(\ \rho^{(m)}_i(\x)\sigtil_j(\y)\beta^{(N)}(\y)
  -\sigtil_i(\x)\beta^{(N)}(\x)\rho^{(m)}_j(\y)\right)
\cr
2K^{\bb}_{MN}[e]&=\int d\x d\y\ K^{(2)}
  [e,\x,\y)^{ij}\left(\sigtil_i(\x)\beta^{(M)}(\x)\sigtil_j(\y)\beta^{(N)}(\y)
  - M\leftrightarrow N\right)\ .
}
\eqno(g)
$$
with $K^{(2)}[e,\x,\y)^{ij}$ defined in \(b).
Similar expressions apply for each $F^{(n)}$.  Furthermore, we
obtain a mode decomposition of the functional derivatives
$\delta F^{(n)}/\delta e(\x)$
by taking the derivative of \(b) with respect to
the dreibein and then expanding $\psi$.  Thus,
  $${{\delta F^{(2)}}\over{\delta e_k(\z)}}[e,r,b]
  =r_{m}r_{m}{{\Delta K^{\rr}_{m n}}\over{\delta e_k(\z)}}[e]+
  r_{m}b_{N}{{\Delta K^{\rb}_{m N}}\over{\delta e_k(\z)}}[e] +
  b_{M}b_{N}{{\Delta K^{\bb}_{M N}}\over{\delta e_k(\z)}}[e]
  \ ,\eqno(h)$$
where
  $${{\Delta K^{\rr}_{mn}}\over{\delta e_k(\z)}}[e]
  ={1\over 2}\int d\x d\y\ {{\delta K^{(2)}[e,\x,\y)^{ij}}\over
  {\delta e_k(\z)}}\left(\rho^{(m)}_i(\x)\rho^{(n)}_j(\y)
  -\rho^{(n)}_i(\x)\rho^{(m)}_j(\y)\right)
  \ ,\eqno(i)$$
and so on.  It is important to specify that we first take the
functional derivative, then expand in modes; since the definition
of the modes depends on the dreibein, these operations do not commute.
The notation $\Delta K^{\rr}_{mn}/\delta e$ serves as a reminder
that this quantity is not equal to $\delta K^{\rr}_{mn}/\delta e$.

Now we may write the constraint \(5.c) in terms of sums over the modes.
A bosonic state will satisfy
$$
\eqalign{
& r_m\left({i\over \sqrt{-h}}
  D_i\rho^{(m)i}(\x)F^{(0)}[e]+{{\hbar\kappa^2}\over 2}
  n_a(\x)\sigma^a\sigbar^b\rho_i^{(m)}(\x)
  {{\delta F^{(0)}[e]}\over{\delta e^b{}_i(\x)}}\right)
  +
\cr
& b_M\left(-2\sqrt{-h}\lambda_M\beta^{(M)} F^{(0)}[e]+{{\hbar\kappa^2}\over 2}
  n_a(\x)\sigma^a\sigbar^b\sigtil_i(\x)\beta^{(M)}(\x)
  {{\delta F^{(0)}[e]}\over{\delta e^b{}_i(\x)}}\right)=0
  \ ,}
\eqno(j)
$$
while the state with Grassmann number 2 satisfies
$$
\eqalign{
& r_m r_n r_l \left({i\over\sqrt{-h}}D_i\rho^{(m)i}(\x) K^{\rr}_{nl}[e]
  +{\hbar\kappa^2\over 2}n_a(\x)\sigma^a\sigbar^b\rho^{(m)}_i(\x)
  {{\Delta K^{\rr}_{nl}}\over{\delta e^b{}_i(\x)}}[e]\right)
\cr
&  +b_M b_N b_L\left(-2\sqrt{-h}\lambda_M\beta^{(M)}(\x) K^{\bb}_{NL}[e]
  +{{\hbar\kappa^2}\over 2}n_a(\x)\sigma^a\sigbar^b\sigtil_i(\x)\beta^{(M)}(\x)
  {{\Delta K^{\bb}_{NL}}\over{\delta e^b{}_i(\x)}}[e]\right)
\cr
&  +r_m r_n b_L \left({i\over\sqrt{-h}}D_i\rho^{(m)i}(\x) K^{\rb}_{nL}[e]
  +{{\hbar\kappa^2}\over 2}n_a(\x)\sigma^a\sigbar^b\rho_i^{(m)}(\x)
  {{\Delta K^{\rb}_{nL}}\over{\delta e^b{}_i(\x)}}[e]
\right.
\cr
& \left. -2\sqrt{-h}\lambda_L\beta^{(L)}(\x) K^{\rr}_{mn}[e]
  +{\hbar\kappa^2\over 2}n_a(\x)\sigma^a\sigbar^b\sigtil_i(\x)\beta^{(L)}(\x)
  {\Delta K^{\rr}_{mn}\over\delta e^b{}_i(\x)}[e]\right)
\cr
&  +r_m b_N b_L \left({i\over\sqrt{-h}}D_i\rho^{(m)i}(\x) K^{\bb}_{NL}[e]
  +{{\hbar\kappa^2}\over 2}n_a(\x)\sigma^a\sigbar^b\rho_i^{(m)}(\x)
  {{\Delta K^{\bb}_{NL}}\over{\delta e^b{}_i(\x)}}[e]
\right.
\cr
& \left. -2\sqrt{-h}\lambda_L\beta^{(L)}(\x) K^{\rb}_{mN}[e]
  +{{\hbar\kappa^2}\over 2}n_a(\x)\sigma^a\sigbar^b\sigtil_i(\x)\beta^{(L)}(\x)
  {{\Delta K^{\rb}_{mN}}\over{\delta e^b{}_i(\x)}}[e]\right)=0\ .
}
\eqno(k)
$$
It is straightforward to obtain the equivalent expression for any
$F^{(n)}$.  Since the coefficients $r_m$ and $b_M$ are arbitrary,
the (appropriately antisymmetrized) expressions in parentheses
must vanish independently.  Our goal is to use this fact to show
that each of the coefficients $K^{\rtob}_{m_1m_2
\ldots M_{n-1}M_n}$ vanish, and hence that each $K^{(n)}$ vanishes.

\subhead{\bf 5.2 The bosonic sector}

We begin by considering the bosonic sector, which we have already
shown in Sec.~2 to have no non-zero solution to the constraint
$\bar{S}^{A'}F^{(0)}=0$, on the basis of a scaling argument. Here we
shall proceed from \(j), in order to introduce a method which generalizes
to states with nonzero Grassmann number.  In \(j), the coefficient of
$b_M$ must be zero (for each value of $M$), or
$$
2\sqrt{-h}\lambda_M\beta^{(M)}(\x) F^{(0)}[e]-{{\hbar\kappa^2}\over 2}
  n_a(\x)\sigma^a\sigbar^b
  \sigtil_i(\x)\beta^{(M)}(\x){{\delta F^{(0)}}[e]\over{\delta e^b{}_i(\x)}}=0
  \ ,
\eqno(l)
$$
where there is no sum on $M$. We shall prove that this implies that
$F^{(0)}$ is identically zero by taking linear (functional) combinations of
this equation for different $M$, in order to eliminate functional
derivatives of $F^{(0)}$. Then we show that the linear combination of the
remaining terms is a non-vanishing function multiplying $F^{(0)}$. From
this we deduce that $F^{(0)}$ vanishes.

The coefficients with which we multiply equation \(l) for different modes
$M$ are encoded in a {\it mode-killing} function $V_M(\x)$, indexed by $M$.
It is enough to consider linear combinations of \(l) for a set of three
modes $\beta^{(M)}(\x)$ in order to define $V_M(\x)$ (although considering
larger sets of modes is permissible). The mode killer is
orthogonal to the $\beta^{(M)}_A(\x)$
(where we have restored the spinor index) for each $A=1,2$ in the sense
that
$$
\sum_{M=1,2,3}\bar V_M(\x)\beta^{(M)}_A(\x)=0\ .
\eqno(m)
$$
Multiplying \(l) for the three modes by the
mode killer serves to remove the term involving $\delta F/
\delta e$, leaving
$$
\left(\sum_{M=1,2,3}\bar V_M(\x) \lambda_M\beta^{(M)}_A(\x)\right)
F^{(0)}[e]=0\ .
\eqno(n)
$$
It is easy to check that the function in brackets does not vanish
identically, as long as the three eigenvalues $\lambda_M$ are
not all equal, and we are free to choose three modes for which this is the
case. Thus, we find that $F^{(0)}$ itself must vanish.

\subhead{\bf 5.3 States with nonzero Grassmann number}

We next turn to states with nonzero Grassmann number.  The procedure
in this case is essentially the same for all $n$, although the
notation quickly becomes unwieldy.  We shall illustrate the method
for $n=2$, and simply outline the extension to larger $n$.  We
begin with the coefficient of $b_Mb_Nb_L$ in \(k); the antisymmetric
part must vanish, which we express as the sum of cyclic permutations:
$$
\eqalign{
& 2\sqrt{-h}\lambda_M
  \beta^{(M)}_A(\x) K^{\bb}_{NL}[e]-{{\hbar\kappa^2}\over 2}
  n_a(\x)\sigma^a\sigbar^b\sigtil_{iA}{}^B(\x)\beta^{(M)}_{B}(\x)
  {{\Delta K^{\bb}_{NL}}\over{\delta e^b{}_i(\x)}}[e]
\cr
&+\{NLM\}+\{LMN\}=0}
  \ .
\eqno(o)
$$
We shall again use mode-killing functions to remove functional derivatives
of $K^{\bb}$. In the case of non-zero Grassmann number states, we
shall find that more than one mode-killing function is required.
It is easy to see, for example, that contracting in $\bar{V}_M(\x)$ will
not make the ${NLM}$ or ${LMN}$ terms with functional derivatives vanish.
If we contract over $N$ using the same $\bar{V}_N(\x)$, then the
antisymmetry implies that the whole of \(o) vanishes. Thus we require enough
modes that we can define (in this case) three different mode killers. The
requisite number of modes is then 5, {\it i.e.} $M=1,\ldots,5$.
(Again, a larger number would be allowed.)
We therefore introduce a new index $\alpha$ labeling different
mode-killing functions: $V^\alpha_M(\x)$. For the sector with
Grassmann number $n$, we require $n+1$ mode killers, $\alpha=1,\ldots,n+1$,
and so $M$ runs over the range $1,\ldots,n+3$.

Returning to the case of Grassmann number 2,
contracting in the three different mode killers, we kill off the terms with
functional derivatives, to obtain
$$
\bar V^{[\alpha}_M(\x)\bar V^\beta_N(\x)\bar V^{\gamma]}_L(\x)
  \left(\lambda_M\beta^{(M)}_A(\x) K^{\bb}_{NL}[e]\right)=0
  \ ,
\eqno(p)
$$
where a sum over $M$, $N$, and $L$ is implied, and square brackets
denote antisymmetrization (the antisymmetrization over $M$, $N$ and
$L$ has been replaced by an equivalent antisymmetrization over $\alpha$,
$\beta$ and $\gamma$).

We now define
$$
U^\alpha_A(\x)=\bar V^\alpha_M(\x)\lambda_M\beta^{(M)}_A(\x)
\eqno(q)
$$
and
$$
G^{\beta\gamma}(\x)=\bar V^{[\beta}_N(\x)\bar V^{\gamma]}_L(\x)
  K^{\bb}_{NL}[e]\ ,
\eqno(r)
$$
which may be thought as a one-form and two-form in a $3$-dimensional
space.  The constraint \(p) can now be
written compactly as the wedge product condition
$$
U^{[\alpha}_A(\x) G^{\beta\gamma]}(\x)=0\ .
\eqno(s)
$$

The wedge orthogonality conditions \(s) imply that
the coefficients $K_{NL}[e]$ must vanish.  To show this we ``rotate'' the
forms $U$ and $G$ by an $\x$-dependent SU($3$) matrix $\Lambda$:
$$
\eqalign{
U^{\alpha^\prime}_A(\x)&=\Lambda^{\alpha^\prime}_\alpha(\x)
  U^\alpha_A(\x)
\cr
  G^{\beta^\prime \gamma^\prime}(\x)&=\Lambda^{\beta^\prime}_\beta(\x)
  \Lambda^{\gamma^\prime}_\gamma(\x) G^{\beta\gamma}(\x)\ ,\cr}
\eqno(u)
$$
choosing $\Lambda$ such that $U^{2^\prime}_1(\x)=U^{3^\prime}_1(\x)=0$.
This leaves
enough freedom in the choice of $\Lambda$ that we may also make
the single component $U^{3^\prime}_2(\x)$ vanish, so that we have
$$
U^{\alpha'}_1(\x)=\Lambda^{\alpha'}_\alpha(\x) U^\alpha_1(\x)
=\pmatrix{\times\cr 0\cr
0},
\qquad
U^{\alpha'}_2(\x)=\Lambda^{\alpha'}_\alpha(\x) U^\alpha_2(\x)
=\pmatrix{\times\cr \times\cr
0}.
\eqno(5.70)
$$
Then Eq.~\(s) implies
$$
G^{2^\prime 3^\prime}(\x)=G^{3^\prime 1^\prime}(\x)=0 \ .
\eqno(v)
$$
We can write these two relations as
$$
G^{2'3'}(\x)\equiv\sum_{N,\ L} Q^{(1)}_{NL}(\x) K^{\bb}_{NL}[e]
  =0,
\qquad
G^{3'1'}(\x)\equiv\sum_{N,\ L} Q^{(2)}_{NL}(\x) K^{\bb}_{NL}[e]
  =0,
\eqno(w)
$$
where
$$
\eqalign{
Q^{(1)}_{NL}(\x)&=\Lambda^{2^\prime}_\beta(\x)\Lambda^{3^\prime}_\gamma(\x)
  \bar{V}^{[\beta}_N(\x) \bar{V}^{\gamma]}_L(\x)\ ,\cr
Q^{(2)}_{NL}(\x)&=\Lambda^{1^\prime}_\beta(\x)\Lambda^{3^\prime}_\gamma(\x)
  \bar{V}^{[\beta}_N(\x) \bar{V}^{\gamma]}_L(\x)\ .\cr}
\eqno(w2)
$$

We now recall from the discussion below \(o) that we are working
with a set of 5 $\beta$-modes $\beta^{(M)}_A(\x)$, $M=1,\ldots 5$.
The 3 associated mode killers $\bar{V}^\alpha_M(\x)$ can be
chosen to satisfy \(m), with non-vanishing components
taking the form
$$
\matrix{
\bar{V}^1(\x)&=&(\bar{V}^1_1(\x),&0,&0,&\bar{V}^1_4(\x),&\bar{V}^1_5(\x))\cr
\bar{V}^2(\x)&=&(0,&\bar{V}^2_2(\x),&0,&\bar{V}^2_4(\x),&\bar{V}^2_5(\x))\cr
\bar{V}^3(\x)&=&(0,&0,&\bar{V}^3_3(\x),&\bar{V}^3_4(\x),&\bar{V}^3_5(\x))
}\ .
\eqno(5.aa)
$$
Each non-zero component is a function of $\x$ which is non-vanishing
in some compact subset $\Sigma^\prime \in \Sigma$.

We now make two plausible technical assumptions which we cannot
fully prove for a general hypersurface $\Sigma$.  We will then
complete the general argument using these assumptions and then
show that the assumptions are valid when $\Sigma$ is flat.
We assume the following:
($a$) The modes $\beta^{(1)}_A(\x)$, $\beta^{(2)}_A(\x)$
are chosen arbitrarily, and the $\beta^{(M)}_A(\x)$, $M=3,4,5$ may
then be chosen so that $Q^{(1)}_{12}(\x)$ is a function which in
$\Sigma^\prime$ is linearly independent of the other $Q^{(1)}_{NL}(\x)$
in \(w), and similarly $Q^{(2)}_{12}(\x)$ is linearly independent
of the other $Q^{(2)}_{NL}(\x)$.  ($b$) All six
components of $U^\alpha_A(\x)$ are nonvanishing in
$\Sigma^\prime$.  This may be arranged by further change in the
modes $\beta^{(M)}_A(\x)$, $M=3,4,5$, if required.  One can show that
this means that no row of the matrix $\Lambda$ can have two
vanishing elements, which will be required below.

Assumption ($a$) means that the terms with coefficient
$K^{\bb}_{12}$ in $G^{2^\prime 3^\prime}(\x)$ and $G^{1^\prime 3^\prime}(\x)$
in \(w) must vanish separately.  We will now show that this can only
happen if $K_{12}=0$.  Since the modes called $\beta^{(1)}_A(\x)$ and
$\beta^{(2)}_A(\x)$ were chosen arbitrarily, all of the $K_{NL}$ must
vanish by similar arguments.  Using \(5.aa), we can write
$$
\eqalign{
Q^{(1)}_{12}(\x)&=(\Lambda^{2^\prime}_1(\x)\Lambda^{3^\prime}_2(\x)
  -\Lambda^{2^\prime}_2(\x)\Lambda^{3^\prime}_1(\x))
  \bar{V}^{1}_1(\x) \bar{V}^{2}_2(\x) \equiv
  \Lambda^{(1)}(\x)\bar{V}^{1}_1(\x) \bar{V}^{2}_2(\x) \ ,\cr
Q^{(2)}_{12}(\x)&=(\Lambda^{1^\prime}_1(\x)\Lambda^{3^\prime}_2(\x)
  -\Lambda^{1^\prime}_2(\x)\Lambda^{3^\prime}_1(\x))
  \bar{V}^{1}_1(\x) \bar{V}^{2}_2(\x) \equiv
  \Lambda^{(2)}(\x)\bar{V}^{1}_1(\x) \bar{V}^{2}_2(\x) \ .\cr}
\eqno(w3)
$$
where $\Lambda^{(1)}(\x)$ and $\Lambda^{(2)}(\x)$
are determinants of
certain $2\times 2$ submatrices of $\Lambda^{\alpha^\prime}_\alpha(\x)$.
Since $\bar{V}^{1}_1(\x) \bar{V}^{2}_2(\x)$ is nonzero by hypothesis,
we need to show that the two determinants $\Lambda^{(1)}(\x)$ and
$\Lambda^{(2)}(\x)$ cannot simultaneously
vanish.  We give a somewhat abstract disscussion, in anticipation
of the higher-$n$ case considered below.  We consider the
first two components of the three rows of
$\Lambda^{\alpha^\prime}_\alpha(\x)$ to define three one-forms $\theta^i(\x)$
in a two-dimensional space:
$$
\Lambda^{\alpha^\prime}_\alpha(\x)=\pmatrix{
\quad&\theta^1(\x) & &\Vrule & . &\cr
\Hrule6\cr
&\theta^2(\x) & &\Vrule & . &\cr
\Hrule6\cr
&\theta^3(\x) & &\LVrule & . &\cr
}
\eqno(5.71)
$$
Thus, the two components of $\theta^1(\x)$ are
$(\Lambda^{1^\prime}_1(\x), \Lambda^{1^\prime}_2(\x))$.  In this notation,
the determinants $\Lambda^{(1)}(\x)$ and $\Lambda^{(2)}(\x)$ will only vanish
if $\theta^2(\x)\wedge\theta^3(\x)=0$ and $\theta^1(\x)\wedge\theta^3(\x)=0$,
respectively; we shall assume this is true and derive a contradiction.
None of the $\theta^i(\x)$ can be identically zero, since
none of the rows of $\Lambda^{\alpha^\prime}_\alpha(\x)$ has two zero
elements.  Therefore we must have $\theta^1(\x)\wedge\theta^2(\x)=0$, and hence
$\Lambda^{(3)}(\x)\equiv\Lambda^{1^\prime}_1(\x)\Lambda^{2^\prime}_2(\x)
-\Lambda^{1^\prime}_2(\x)\Lambda^{2^\prime}_1(\x)=0$.  This in turn implies
that Det$~\Lambda=0$, which is impossible for $\Lambda\in$ SU(3).
Thus, either $\Lambda^{(1)}(\x)$ or $\Lambda^{(2)}(\x)$ must be nonzero,
and hence $K_{12}$ must vanish in order to satisfy \(w).  Since
there were no special properties of the specific modes $\beta^{(1)}(\x)$ and
$\beta^{(2)}(\x)$ used in this demonstration, all of the $K^{\bb}
_{NL}$ coefficients will vanish by similar arguments.

Having shown that $K_{NL}=0$
subject to assumptions ($a$) and ($b$) as stated below \(5.aa),
we now demonstrate that these assumptions hold in flat space.
In this case the modes $\beta^{(M)}(\x)$ are constant spinors times
exp$(i\k_M\cdot\x)$.  The mode killers may therefore be chosen
to be of the form
$$
\bar{V}^\alpha_M(\x)=A^\alpha_M e^{-i\k_M\cdot\x}\ ,\eqno(5.na)
$$
where the $A^\alpha_M$ are constants, vanishing and not vanishing
as in \(5.aa).  Then \(w) takes the form
$$
\eqalign{
& G^{2'3'}(\x)\equiv\sum_{N,\ L} R^{(1)}_{NL} K^{\bb}_{NL}[e]
  e^{-i(\k_N+\k_L)\cdot\x}=0,
\cr
& G^{3'1'}(\x)\equiv\sum_{N,\ L} R^{(2)}_{NL} K^{\bb}_{NL}[e]
  e^{-i(\k_N+\k_L)\cdot\x}=0 \ ,}
\eqno(www)
$$
with $R^{(1)}_{NL}$ and $R^{(1)}_{NL}$ constant.  With modes
$\beta^{(1)}(\x)$ and $\beta^{(2)}(\x)$ chosen arbitrarily, it is clear
that we can arrange $\beta^{(M)}(\x)$ for $M=3,4,5$ so that the
linear independence assumed in ($a$) is satisfied.  Furthermore
$U^\alpha_A$ is constant, and it is easy to guarantee that
assumption ($b$) is satisfied everywhere on $\Sigma$.
The proof that all the $K_{NL}$ must vanish
is thus complete for flat $\Sigma$, but rests on the
assumptions ($a$) and ($b$) which are not proven for general
$\Sigma$.

Using the other terms in the constraint \(k), it is straightforward
to show that $K^{\rb}_{mN}$ and $K^{\rr}_{MN}$
must also vanish.  Consider the mixed term
$K^{\rb}_{mN}$.  Since we have already shown that
$K^{\bb}_{NL}=0$ for all $L$ and $N$, setting the last term in \(k) equal
to zero implies
$$
\eqalign{
&-{{\hbar\kappa^2}\over 2}n_a(\x)\sigma^a\sigbar^b\rho_i^{(m)}(\x)
  {{\Delta K^{\bb}_{NL}}\over{\delta e^b{}_i(\x)}}[e]
  +2\sqrt{-h}\lambda_L\beta^{(L)}(\x) K^{\rb}_{mN}[e]
\cr
&-{{\hbar\kappa^2}\over 2}n_a(\x)\sigma^a\sigbar^b\sigtil_i(\x)\beta^{(L)}(\x)
  {{\Delta K^{\rb}_{mN}}\over{\delta e^b{}_i(\x)}}[e]
  -\{N\leftrightarrow L\}=0\ .
}
\eqno(aa)
$$
As before, we can get rid of the term containing $\sigtil_i\beta^{(L)}(\x)$
by multiplying by mode killers $\bar V^\alpha_L(\x)$. In this case we require
two mode killers so that $\alpha=1,2$ and $L$ and $N$ run over 1 to 4.
Likewise, we can introduce a new set of mode-killing
functions $W_m(\x)$ which are orthogonal to the $\rho^{(m)}_i(\x)$
modes:
$$
\sum_m\bar W_m(\x)\rho^{(m)}_i(\x)=0\ .
\eqno(ab)
$$
In this case we require only one of these mode-killers, and since
$\rho^{(m)i}_A(\x)$ has six components, we need 7 distinct modes, {\it i.e.}
$m=1,\ldots,7$.
Multiplying \(aa) by these
mode killers, only the second term in \(aa) survives, so that
$$
\bar{V}^{[\alpha}_L(\x)\bar{V}^{\beta]}_N(\x)
\bar{W}_m(\x) \lambda_L\beta^{(L)}(\x)K^{\rb}_{mN}[e]=0\ .
\eqno(5.ab)
$$
We may rewrite this as
$$
U^{[\alpha}_A(\x) H^{\beta]}(\x)=0\ ,
\eqno(ac)
$$
where
$$
U^\alpha_A(\x)=\bar V^\alpha_L(\x) \lambda_L\beta^{(L)}_A(\x)\ ,
\eqno(ad)
$$
and we have introduced
$$
H^\beta(\x)=\bar W_m(\x)\bar V^\beta_N(\x) K^{\rb}_{mN}[e]\ .
\eqno(ae)
$$
Although \(ac) is not of precisely the same form as \(s),
the same reasoning applies (with SU(2) rather than SU(3)), and
suffices to show that the mixed coefficients $K^{\rb}_{mN}$
must all vanish.

Finally, the $K^{\rr}_{mn}$ part of the wave function can be
eliminated using the first term in \(k).  Defining a set of three
mode killers $\bar W^\alpha_m(\x)$ which satisfy \(ab), we once again
obtain an equation of the form \(s), this time with
$$
\eqalign{
U^\alpha_A(\x)&=\bar W^\alpha_m(\x) D_i\rho^{(m)i}_A(\x)\ ,
\cr
G^{\beta\gamma}(\x)&=\bar W^{[\beta}_n(\x)\bar W^{\gamma]}_l(\x)
  K^{\rr}_{nl}[e]\ .}
\eqno(af)
$$
{}From here the analysis is almost identical to that following
\(s).  The single exception arises due to the existence of
modes with $D^i\rho^{(m)}_i(\x)=0$, which do not contribute to the
definition of $U^\alpha_A(\x)$.  However, it is possible to consider
sets of modes for which we obtain a nonzero $D^i\rho^{(m)}_i(\x)$
multiplying any desired term $K^{\rr}_{mn}$, which is
sufficient to show that every $K^{\rr}_{mn}$ must
vanish, as can be easily checked.

This completes the demonstration that there are no
states with Grassmann number 2.  The generalization to higher
Grassmann numbers follows the above procedure very closely, so
we will just outline the major steps.
In a Grassmann number $n$ state, the part of the constraint which
involves only $\beta$ modes leads to
$$
U^{[\alpha}_A(\x)G^{\alpha_1 \alpha_2\ldots\alpha_n]}(\x)=0
  \ ,
\eqno(ag)
$$
where $U^\alpha_A(\x)$ is defined in \(q) and
$$
G^{\alpha_1 \alpha_2\ldots\alpha_n}(\x)=\sum_{M_1\ldots M_n}
  \bar V^{[\alpha_1}_{M_1}(\x)\ldots\bar V^{\alpha_n]}_{M_n}(\x)
  K^{\btob}_{M_1\ldots M_n}[e]
\eqno(ah)
$$
is the natural generalization of \(r).  We rotate $U^\alpha_A(\x)$ and
$G^{\alpha_1 \alpha_2\ldots\alpha_n}(\x)$ by SU($n+1$) matrices
$\Lambda^{\alpha^\prime}_\alpha(\x)$, such that only the first one
component of $U^{\alpha^\prime}_1(\x)$ and the first two components of
$U^{\alpha^\prime}_2(\x)$ are nonvanishing.  (This requirement leaves
a residual SU($n-1$) freedom, which we will take advantage of
below.)  By judicious choice of the mode killers,
the constraint equation implies that
either $K^{\btob}_{1\ldots n}=0$, or both of the determinants
$$
\eqalign{
  \Lambda^{(1)}(\x)&\equiv \Lambda^{2^\prime}_{[1}(\x)
\Lambda^{3^\prime}_{2}(\x)
  \ldots \Lambda^{(n+1)^\prime}_{n]}(\x)\ ,\cr
  \Lambda^{(2)}(\x)&\equiv \Lambda^{1^\prime}_{[1}(\x)
\Lambda^{3^\prime}_{2}(\x)
  \ldots \Lambda^{(n+1)^\prime}_{n]}(\x)\ ,\cr}
\eqno(bh)
$$
must vanish.

As before, we define a set of $n+1$ one-forms $\theta^i(\x)$ in an
$n$-dimensional space, where the components of $\theta^i(\x)$ are the
first $n$ elements of row $i$ of $\Lambda^{\alpha^\prime}_\alpha(\x)$:
$$
\Lambda^{\alpha^\prime}_\alpha(\x)=\pmatrix{
\qquad &\theta^1(\x) &\quad&\Vrule & . &\cr
\Hrule6\cr
&\theta^2(\x)&&\Vrule & .&\cr
\Hrule6\cr
&&&\Vrule&&\cr
&\vdots&&\Vrule & \vdots &\cr
&&&\Vrule&&\cr
\Hrule6\cr
&\theta^{n+1}(\x)&&\LVrule & .& \cr
}
\eqno(5.72)
$$
The hypothesis that both $\Lambda^{(1)}(\x)$ and $\Lambda^{(2)}(\x)$
vanish is equivalent to
$$
\eqalign{
  \theta^2(\x)\wedge\theta^3(\x)\ldots \theta^{n+1}(\x)=0\ ,\cr
  \theta^1(\x)\wedge\theta^3(\x)\ldots \theta^{n+1}(\x)=0\ .\cr}
\eqno(bi)
$$
None of the $\theta^i(\x)$ have all zero components; hence, we have either
that $\theta^1(\x)\wedge\theta^2(\x)=0$, or another pair of forms are
wedge-orthogonal: $\theta^j(\x)\wedge\theta^k(\x)=0$ for some $j,k >2$.
In the former case, all of the determinants $\Lambda^{(i)}(\x)$ must
vanish, and hence Det$~\Lambda$ itself will vanish, thus yielding
a contradiction.  In the latter case, the residual SU($n-1$)
freedom left over after the values of $U^{\alpha^\prime}_A(\x)$ have
been fixed may be used to rotate the components of $\theta^i(\x)$
and $\theta^j(\x)$ into one another such that one of these forms
vanishes identically; however, since none of the forms may
vanish this is also a contradiction.  Thus one of the two
determinants $\Lambda^{(1)}(\x)$, $\Lambda^{(2)}(\x)$ must be nonzero, and
the constraint can only be satisfied if
$K^{\btob}_{1\ldots n}$ vanishes.  By similar arguments all the
other $K^{\btob}_{M_1\ldots M_n}$, as well as the terms
involving the $\rho$ modes, suffer the same fate.

\head{\bf 6. Conclusions}
\taghead{6.}

The main conclusion of this paper is that there are no physical states in
$N=1$ supergravity that are purely bosonic or have fixed finite Grassmann
number $n$, nor can there be any state which is a linear superposition of
components with finite $n$. We conclude that all
physical states in $N=1$ supergravity must have infinite Grassmann number.
It is curious that these results
emerge from the supersymmetry constraints whose form is consistent with an
expansion in states of definite Grassmann number.
A study of the free spin-3/2 field gives a similar structure for physical
(gauge invariant) states
in that theory, and the infinite Grassmann number is seen to
arise from the essentially multiplicative character of the gauge constraint
\(4.5). The constraint is
satisfied by including a delta functional over an
infinite number of fermionic modes
in the state functional, realized as an infinite product of
anti-commuting fields.
In the interacting theory, the $\bar{S}$ constraint can again be
made to act multiplicatively after projection with mode-killing functions
in a similar way to \(4.5) and \(4.28),
accounting for the absence of finite Grassmann
number physical states.

The similarity between the results for the free and interacting theories
can be contrasted with the situation in non-abelian gauge theories. The
most apt comparison is with the electric-field formulation of SU(2) gauge
theory of Goldstone and Jackiw [\cite{GJ}]. There the Gauss law constraint
for a free gauge field forces the physical state to contain an infinite
product of delta functions in the longitudinal modes $k^iE^a{}_i(\x)$. The
states of the interacting theory, on the other hand, may be arbitrary
functionals of the gauge-invariant tensor $S_{ij}=E^a{}_i(\x)E^a{}_j(\x)$.
In supergravity we know of no local gauge invariant variables analogous to
$S_{ij}$, while the fact that infinite Grassmann number can be established
for the interacting theory as well as the free theory indicates that
physical states in the full theory are closer to the free-field form than
in non-abelian gauge theories.

Our results on the structure of physical states may also be contrasted with
previous work on minisuperspace models of $N=1$ supergravity [\cite{mini}]
in which purely bosonic physical states were found. The reason for this
apparent discrepancy is the severe restriction on the allowed spatial
dependence in minisuperspace models, and the consequent simplification of
the constraints \(2.100)-\(2.101). In most cases the gravitino derivative
terms are completely absent from the constraints, and so bosonic solutions
are allowed for the same reason that the zero modes of the free gravitino
field are not restricted by the constraints in \(4.28)-\(4.29).

It is our hope that the results derived here have clarified the issues
raised in [\cite{d93}] and, independently, have illuminated the structure
of physical states in canonically quantized
supergravity. We must express the reservation that the
procedure of solving the formal constraint equations without due
attention to possible quantum anomalies
is problematic [\cite{nicolai2}], and we
hope that the issue of anomalies is addressed in future work. The
apparently simpler constraint structure of supergravity, when compared
to that of general relativity, makes us optimistic about further
progress.

\head{\bf Acknowledgments}

We wish to thank Andrei Barvinsky, Nathan Berkovits,
Roger Brooks, Bernard de~Wit, Peter D'Eath, Gary Gibbons,
Marc Henneaux, Viqar Husain, Roman Jackiw, Hans-J\"urgen Matschull,
Hermann Nicolai and Barton Zwiebach for
discussions which have influenced our work.  This work has been
supported in part by the U.S. National Science Foundation under
grant PHY/92-06867, by the U.S. Department of Energy under
contract DE-AC02-76ER03069, and by the Natural Sciences and
Engineering Research Council of Canada.

\head{\bf Appendix: Notation and Some Useful Identities}
\taghead{A.}

We have chosen the following definitions for the Weyl spinor formalism:
$$
\eqalign{
& \ve_{AB}=\pmatrix{0 & -1\cr 1 & 0},\qquad
\ve^{AB}=\pmatrix{0 & 1\cr -1 & 0}
\cr
& \ve^{AB}\phi_B=\phi^A,\qquad\ve_{AB}\phi^B=\phi_A
\cr
& {\ve_A}^B=-{\delta_A}^B,
\cr
& \sigma^a_{AA'} =  (1,\sigma_i)
\cr
& \eta_{ab}\sigma^a_{AA'}\sigma^b_{BB'}=2\ve_{AB}\ve_{A'B'}
\cr
& {\sigbar_a}^{A'A} =  \ve^{AB}\ve^{A'B'}\sigma_{aBB'} =  (1,\sigma_i)\ .
}
\eqno(A.1)
$$
The sigma matrices defined in this way obey a series of identities which
are essential in deriving some of the results given in Secs. 2 and 4:
$$
\eqalign{
& \sigma^a\sigbar^b-\sigma^b\sigbar^a=-i{\ve^{ab}}_{cd}\sigma^c\sigbar^d
\cr
& \sigbar^a\sigma^b-\sigbar^b\sigma^a=i{\ve^{ab}}_{cd}\sigbar^c\sigma^d
\cr
& \sigma^a_{AA'}\sigbar^{bB'B}-\sigma^b_{AA'}\sigbar^{aB'B}=
{1\over 2}\left[{(\sigbar^b\sigma^a)^{B'}}_{A'}{\delta_A}^B
+{(\sigma^a\sigbar^b-\sigma^b\sigbar^a)_A}^B{\delta^{B'}}_{A'}
\right]
\cr
& \sigma^a\sigbar^b+\sigma^b\sigbar^a=2\eta^{ab}
\cr
& \sigbar^a\sigma^b+\sigbar^b\sigma^a=2\eta^{ab}
\cr
& \sigma^a\sigbar^b\sigma^c+\sigma^c\sigbar^b
\sigma^a=2\left(\eta^{ab}\sigma^c
+\eta^{bc}\sigma^a-\eta^{ac}\sigma^b\right)
\cr
& \sigbar^a\sigma^b\sigbar^c+\sigbar^c\sigma^b\sigbar^a=
2\left(\eta^{ab}\sigbar^c+\eta^{bc}\sigbar^a-\eta^{ac}\sigbar^b\right)
\cr
& \sigma^a\sigbar^b\sigma^c-\sigma^c\sigbar^b
\sigma^a=2i\ve^{abcd}\sigma_d
\cr
& \sigbar^a\sigma^b\sigbar^c-\sigbar^c\sigma^b\sigbar^a
=-2i\ve^{abcd}\sigbar_d
\cr
& e_{ai}e_{bj}\sigma^a\sigbar^b=h_{ij}-i\ve_{ijk}n^ce^{dk}\sigma_c
\sigbar_d\sqrt{-h}
\cr
& e_{ai}e_{bj}\sigbar^a\sigma^b=h_{ij}+i\ve_{ijk}n^ce^{dk}\sigbar_c
\sigma_d\sqrt{-h}
\cr
& n^an^b(\sigma_a\sigbar_b)_A{}^B=\delta_A{}^B,\qquad
n^an^b(\sigbar_a\sigma_b)^{A'}{}_{B'}=\delta^{A'}{}_{B'}\ .
}
\eqno(A.2)
$$
Finally, we have defined
$$
\eqalign{
& \ve^{0123}=\ve^{123}=1
\cr
& \ve^{ijk}=\ve^{abcd}n_a{e_b}^i{e_c}^j{e_d}^k\sqrt{-h}
\cr
& \ve_{ijk}=\ve^{abcd}n_ae_{bi}e_{cj}e_{dk}{1\over\sqrt{-h}}\ ,
}
\eqno(A.3)
$$
so that $\ve_{ijk}$ is a tensor density.

\endpage

\references

\refis{d84} P. D. D'Eath, Phys. Rev. D {\bf 29} (1984) 2199.

\refis{pilati} M. Pilati, Nucl. Phys. {\bf B132} (1978) 138.

\refis{teitelboim} C. Teitelboim, Phys. Rev. Lett. {\bf 38} (1977) 1106.

\refis{fradkin} E. S. Fradkin and M. A. Vasiliev, Phys. Lett. {\bf 72B}
(1977) 70.

\refis{d93} P. D. D'Eath, ``Physical states in $N=1$ supergravity", DAMTP
preprint, hep-th/9304084.

\refis{dandas} A. Das and D. Z. Freedman, Nucl. Phys. {\bf B114} (1976) 271.

\refis{mini} R. Graham, Phys. Rev. Lett. {\bf 67} (1991) 1381;
P. D. D'Eath and D. I. Hughes, Nucl. Phys. {\bf B378} (1992) 381;
P. D. D'Eath, S. W. Hawking and O. Obreg\'on, Phys. Lett. {\bf B300}
(1993) 44; P. D. D'Eath, Phys. Rev. D {\bf 48} (1993) 713;
R. Graham and H. Luckock, ``The Hartle-Hawking state for the
Bianchi IX model in supergravity'', gr-qc/9311004.

\refis{nicolai2} H.-J. Matschull and H. Nicolai, ``Canonical Quantum
Supergravity in Three Dimensions,'' gr-qc/9306018.

\refis{nicolai} B. de Wit, H.-J. Matschull and H. Nicolai, Phys. Lett.
{\bf 318B} (1993) 115.

\refis{cs} M. Henneaux and  C. Teitelboim, {\it Quantization of
Gauge Systems} (Princeton University Press, Princeton, 1992).

\refis{fadslav} L.D. Faddeev and A.A. Slavnov, {\it Gauge Fields: An
Introduction to Quantum Theory}, 2nd ed. (Addison Wesley, Redwood
City, California, 1991).

\refis{GJ} J. Goldstone and R. Jackiw, Phys. Lett. {\bf 74 B} (1978) 81.

\refis{senja} G. Senjanovich, Phys. Rev. D {\bf 16} (1977) 307.

\endreferences

\end